\documentclass[prl,aps,showpacs,byrevtex,twocolumn,groupedaddress]{revtex4}
\usepackage{amsfonts}
\usepackage{graphicx}
\usepackage{amsmath}

\input{tcilatex}

\begin{document}

\title{Semiclassical quantization of electrons in magnetic fields: the
generalized Peierls substitution }
\author{Pierre Gosselin$^{1}$, Hocine Boumrar$^{2,3}$ and Herv\'{e} Mohrbach$%
^{2}$}
\affiliation{$^{1}$Institut Fourier, UMR 5582 CNRS-UJF, UFR de Math\'{e}matiques,
Universit\'{e} Grenoble I, BP74, 38402 Saint Martin d'H\`{e}res, Cedex,
France }
\affiliation{$^{2}$Laboratoire de Physique Mol\'{e}culaire et des Collisions, ICPMB-FR
CNRS 2843, Universit\'{e} Paul Verlaine-Metz, 57078 Metz Cedex 3, France}
\affiliation{$^{3}$Laboratoire de Physique et Chimie Quantique, Universit\'{e} Mouloud
Mammeri -BP 17, Tizi Ouzou, Algerie}

\begin{abstract}
A generalized Peierls substitution which takes into account a Berry phase
term must be considered for the semiclassical treatment of electrons in a
magnetic field. This substitution turns out to be an essential element for
the correct determination of the semiclassical equations of motion as well
as for the semiclassical Bohr-Sommerfeld quantization condition for energy
levels. A general expression for the cross-sectional area is derived and
used as an illustration for the calculation of energy levels of Bloch and
Dirac electrons.
\end{abstract}

\maketitle

Semiclassical approaches are very important in many area of physics for the
study of the short wave length behavior of quantum systems, including Bloch
electrons in crystals or Dirac particles in external fields. An essential
ingredient of these approaches is the Bohr-Sommerfeld quantization
condition, whose generalization from scalar to vector wave fields has
revealed new gauge structures related to Berry's phases \cite{LITTLEJOHN}.
This paper presents a detailed study of the semiclassical quantization for a
single quantum particle in a magnetic field, exemplified by electrons in a
crystal and by Dirac electrons. This unified description of a particle in a
magnetic field is based on a method of semiclassical diagonalization for an
arbitrary matrix valued Hamiltonian developed previously \cite{PIERRESC}
(for a generalization to higher order in $\hbar $ see \cite{PIERRE}). This
method results in an effective diagonal Hamiltonian in terms of
gauge-covariant but noncanonical, actually noncommutative, coordinates. It
will be shown that a generalized Berry's phase dependent Peierls
substitution is necessary for the establishment of the full equations of
motion including Berry's phase terms. This substitution turns out to be also
an essential ingredient for the Bohr-Sommerfeld quantization condition of an
electron in a magnetic field. Indeed, when reformulated in terms of the
generalized Peierls substitution, this condition leads to a modification of
the semiclassical quantization rules as well as to a generalization of the
cross-sectional area derived independently by Roth \cite{Roth} and
Fal'kovskii \cite{KALKOVSKII} in the context of Bloch electrons.

\textit{Semiclassical diagonalization. }Let us consider a system of a
quantum particle in an uniform external magnetic field $\ \mathbf{B}=\mathbf{%
\nabla }\times \widetilde{\mathbf{A}}$ described by an arbitrary matrix
valued Hamiltonian $H\left( \mathbf{\Pi ,R}\right) $, where $\mathbf{\Pi }=%
\mathbf{P}+e\widetilde{\mathbf{A}}\mathbf{(R)}$ is the covariant momentum
and $e>0$ is the electric charge. We assume that the system can be separated
into two contributions such that $H\left( \mathbf{\Pi ,R}\right)
=H_{m}\left( \mathbf{\Pi }\right) +\varphi (\mathbf{R})$ where $H_{m}\left( 
\mathbf{\Pi }\right) $ is the pure magnetic part and $\varphi (\mathbf{R})$
is the external electric potential. In this paper we will be mainly
interested by the magnetic contribution. The exact diagonalization of this
matrix valued operator through an unitary matrix $U\left( \mathbf{\Pi }%
\right) $ is in general not known, and in this paper we apply a recursive
diagonalization procedure developed previously by two of the authors. This
procedure is based on a series expansion in the Planck constant of the
required diagonal Hamiltonian \cite{PIERRESC}\cite{PIERRE}. By diagonal
Hamiltonian it is meant a matrix representation with block-diagonal matrix
elements associated with energy band subspaces\textbf{.} The method is based
on the knowledge of the zero-order diagonal representation $\varepsilon
=U_{0}H_{0}U_{0}^{+}$ where $U_{0}$ is the zero-order transformation matrix, 
$H_{0}$ the Hamiltonian $H_{m},$ in which the components $\Pi ^{i}$ are
considered formally as classical, and therefore commuting operators. Quantum
corrections are then re-introduced to yield the expression for the diagonal
Hamiltonian $H_{d}=U\left( \mathbf{\Pi }\right) HU^{+}\left( \mathbf{\Pi }%
\right) $ which, if we limit ourselves to the semiclassical order (the
semiclassical condition being that the radius of curvature of the orbit is
large in comparison with wavelength), has diagonal operator elements
labelled by the energy index $n$ which reads : 
\begin{equation}
\left( H_{d}\right) _{nn}=\varepsilon _{n}\left( \mathbf{\pi }_{n}\right)
+\varphi (\mathbf{r}_{n})-e\hbar \mathbf{M.B}  \label{H}
\end{equation}%
Here $\varepsilon _{n}\left( \mathbf{\pi }\right) $ is the zero-order matrix
element of $\varepsilon $ (it can itself be a matrix as for a Dirac
Hamiltonian, in which case a block-diagonalization is considered) in which
classical variables are now replaced by the quantum operators $\mathbf{\pi }%
_{n}=\mathbf{\Pi }+\hbar \emph{A}_{\pi }$ and $\mathbf{r}_{n}=\mathbf{R}%
+\hbar \emph{A}_{r}$, where we have defined the Berry connection as being
the projection on the $n$th energy band $\emph{A}_{r/\pi }=\mathcal{P}_{n}%
\left[ \mathcal{A}_{r/\pi }\right] $ of the matrix $\mathcal{A}_{r/\pi }=\pm
i\left[ U\mathbf{\nabla }_{\pi /r}U^{+}\right] $. It turns out that in the
magnetic case the Berry connection in momentum has the following expression $%
\emph{A}_{\pi }=-e\emph{A}_{r}\times \mathbf{B}$. Looking at Eq. $\left( \ref%
{H}\right) $ we have proven that instead of the Peierls substitution \cite%
{PEIERLS}, which amounts to replace the canonical momentum $\mathbf{P}$ by
the covariant one $\mathbf{\Pi }$ in the energy band $\varepsilon _{n}$, one
has to consider a generalization of the Peierls substitution via the
noncanonical covariant momentum 
\begin{equation}
\mathbf{\pi }_{n}=\mathbf{\Pi }-e\hbar \emph{A}\times \mathbf{B}
\label{pinc}
\end{equation}%
(where we now use the notation $\emph{A}\equiv \emph{A}_{r}$)$\mathbf{.}$
The last term in Eq. $\left( \ref{H}\right) $ is the coupling between the
uniform magnetic field and the magnetic moment defined as $\mathbf{M}\left(
\pi \right) =\frac{i}{2}\mathcal{P}_{n}\left( \left[ \varepsilon ,\mathcal{A}%
\right] \times \mathcal{A}\right) =\frac{\hbar }{2}\mathcal{P}_{n}\left( 
\overset{\cdot }{\mathcal{A}}\times \mathcal{A}\right) .$ Note that it is
common (especially in solid state physics \cite{LANDAU}) to write the matrix
elements of the components of $\mathbf{M}$ as $M^{i}=\frac{i\hbar
^{2}\varepsilon ^{ijk}}{2}\sum\limits_{m\neq n}\frac{(\overset{\cdot }{%
\mathcal{A}}_{j})_{nm}(\overset{\cdot }{\mathcal{A}}_{k})_{mn}}{\varepsilon
_{n}-\varepsilon _{m}}$ (where we used $\overset{\cdot }{\mathcal{A}}_{nm}=%
\frac{i}{\hbar }\left( \varepsilon _{n}-\varepsilon _{m}\right) \mathcal{A}%
_{nm}$), which thus depend on the band-to-band matrix element of $\mathcal{A}
$.

The appearance of the Berry connection allows us to define naturally
non-Abelian (in general) Berry curvatures $\Theta _{ij}\left( \mathbf{\pi }%
\right) =\partial _{r^{i}}\emph{A}_{j}-\partial _{r^{j}}\emph{A}_{i}+\left[ 
\emph{A}_{_{i}},\emph{A}_{_{j}}\right] $ where for simplicity we omit now
band indices. Position operators then satisfy an unusual non-commutative
algebra $\left[ r^{i},r^{j}\right] =i\hbar ^{2}\Theta ^{ij}$. The
generalized covariant momentum satisfy an algebra $\left[ \pi ^{i},\pi ^{j}%
\right] =-ie\hbar \varepsilon ^{ijk}B_{k}+ie^{2}\hbar ^{2}\varepsilon
^{ipk}\varepsilon ^{jql}\Theta ^{pq}B_{k}B_{l}$ slightly corrected with
respect to the usual one $\left[ \Pi ^{i},\Pi ^{j}\right] =-ie\hbar
\varepsilon ^{ijk}B_{k}$ by a term of order $\hbar ^{2}B^{2}$ which can in
general be neglected. The Heisenberg relations between the coordinate and
the momentum $\left[ r^{i},\pi ^{j}\right] =i\hbar \delta ^{ij}+ie\hbar
^{2}\varepsilon ^{jlk}\Theta ^{il}B_{k}$ is also slightly changed but by a
term of order $\hbar ^{2}B$. This contribution which is a direct consequence
of introducing the generalized covariant momentum was overlooked in previous
works, with the exception of Bliokh's work on the specific case of the Dirac
equation \cite{BLIOKH}. It turns out that this term is essential for the
determination of the genuine semiclassical equations of motion which are 
\begin{eqnarray}
\overset{\cdot }{\mathbf{r}} &=&\partial \mathcal{E}/\partial \mathbf{\pi }%
-\hbar \dot{\mathbf{\pi }}\times \Theta (\mathbf{\pi })  \notag \\
\dot{\mathbf{\pi }} &=&-e\mathbf{E}-e\dot{\mathbf{r}}\times \mathbf{B}
\label{EQM}
\end{eqnarray}%
where we defined $\mathcal{E}\equiv \varepsilon -e\hbar \mathbf{M.B}$. As
consequence of the non-commutative algebra, the velocity equation is
corrected by an anomalous velocity term $\dot{\mathbf{\pi }}\times \Theta $,
where the vector $\Theta $ defined as $\Theta _{i}=\varepsilon _{ijk}\Theta
^{jk}/2$ is the Berry curvature of an electronic state in the given $n$th
band, associated to the electron motion in the $n$th energy band. These
equations of motion, where first derived in solid state physics context in 
\cite{NIU} (see also \cite{PIERRESOLIDE}) by considering the evolution of
the wave packet of a Bloch electron in an electromagnetic field. In this
picture, it is the mean over wave packets of the operator $\mathbf{r}$
corresponding thus to the wave-packet center $r_{c}$ and the mean of $%
\mathbf{\pi }$ giving the mean wave vector $\pi _{c}$ that are the variables
in Eq $\left( \ref{EQM}\right) $. The operatorial approach reveals first
that the operator $\mathbf{\pi }$ is in fact a generalized covariant
momentum operator which replaces the Peierls substitution, and second, that
the operatorial equations of motion are not restricted to Bloch electrons in
a magnetic field but are valid for any physical system described by an
arbitrary matrix valued Hamiltonian\ of the kind $H\left( \mathbf{\Pi ,R}%
\right) =H_{m}\left( \mathbf{\Pi }\right) +\varphi (\mathbf{R}).$ In
particular they are also valid for Dirac particles moving in an
electromagnetic field.

Note that in solids, for crystals with simultaneous time-reversal and
spatial inversion symmetry, the Berry curvature and the magnetic moment
vanish identically throughout the Brillouin zone \cite{NIU}. This is the
case for most applications in solid state physics, but there are situations
where these symmetries are not simultaneously present as in GaAs, where
inversion symmetry is broken, or in ferromagnets, which break time reversal
symmetries. In the same way, the presence of a strong magnetic field, the
magnetic Bloch bands corresponding to the unperturbated system breaks the
time inversion symmetries. In all these cases the dynamical and transport
properties must be described by the full equations of motion given by Eq. $%
\left( \ref{EQM}\right) $. In the case of Dirac particles, both the Berry
curvature and the magnetic moment are non zero and the full equations of
motion have to be considered \cite{PIERRESC}\cite{BLIOKH}.

\textit{Bohr-Sommerfeld quantization. }Having shown the necessity\textbf{\ }%
of the generalized Peierls substitution for the determination of the
semiclassical equations of motion, we now investigate the relevance of this
new concept at the level of the semiclassical quantization of the energy
levels for an electron motion in an external uniform magnetic field only $%
\left( \varphi =0\right) $, so that Eq. $\left( \ref{EQM}\right) $ becomes $%
\mathbf{\dot{r}}=D\left( \frac{\partial \mathcal{E}}{\partial \mathbf{\pi }}%
\right) $\ and \ $\overset{\cdot }{\mathbf{\pi }}=-eD\left( \frac{\partial 
\mathcal{E}}{\partial \mathbf{\pi }}\mathbf{\times B}\right) $ with $%
D^{-1}=1-e\hbar \mathbf{B\Theta .}$ For convenience, $\mathbf{B}$ is chosen
to point in the $z$-direction $\mathbf{B=}B\mathbf{k}.$ Consequently the
orbits satisfies the conditions $\mathcal{E}=$const and $\pi _{z}=$const.
The semiclassical quantization of energy levels can be done according to the
Bohr-Sommerfeld quantization rule 
\begin{equation}
\oint P_{\perp }dR_{\perp }=2\mathcal{\pi }\hbar \left( n+1/2\right)
\label{BS}
\end{equation}%
where $P_{\perp }$ is the canonical momentum in the plane perpendicular to
the axis $\pi _{z}=cte.$ The integration is taken over a period of the
motion and $n$ is a large integer. Now, it turns out to be convenient to
choose the gauge $\widetilde{A}_{y}=BX,$\ $\widetilde{A}_{x}=\widetilde{A}%
_{z}=0.$ In this gauge, one has $\pi _{z}=P_{z}=cte$, and the usual
covariant momentum $\Pi _{y}=P_{y}+eBX.$ As $BX=B(x-\hbar \emph{A}_{x})$ the
generalized covariant momentum defined as $\pi _{y}=\Pi _{y}+e\hbar B\emph{A}%
_{x}$ becomes 
\begin{equation}
\pi _{y}=P_{y}+eBx  \label{piy}
\end{equation}%
which is formally the same relation as the one between the canonical
variables, but now relating the new covariant generalized dynamical
operators. This relation with the help of the equations of motion gives $%
\overset{\cdot }{P}_{y}=\overset{\cdot }{\pi }_{y}-eB\overset{\cdot }{x}=0$
thus $P_{y}$ is a constant of motion so that $\oint P_{y}dY=P_{y}\oint dY=0$
and Eq.$\left( \ref{BS}\right) $ becomes simply $\oint P_{x}dX=2\mathcal{\pi 
}\hbar \left( n+1/2\right) .$ Now using the definition of the generalized
momentum $P_{x}=\pi _{x}+e\hbar \emph{A}_{y}B$ and the differential of the
canonical position $dX=dx-\hbar d\emph{A}_{x}=\frac{d\pi _{y}}{eB}-\hbar d%
\emph{A}_{x},$ the Bohr-Sommerfeld condition Eq. $\left( \ref{BS}\right) $
becomes 
\begin{equation}
\oint \pi _{x}d\pi _{y}=2\mathcal{\pi }\hbar eB\left( n+\frac{1}{2}-\frac{1}{%
2\mathcal{\pi }}\oint \emph{A}_{\bot }d\mathbf{\pi }_{\perp }\right)
\label{pipi}
\end{equation}%
where the integral is now taken along a closed trajectory $\Gamma $ in the $%
\pi $ space and $\frac{1}{2\mathcal{\pi }}\oint \emph{A}_{\bot }d\mathbf{\pi 
}_{\perp }=\phi _{B}$ is the Berry phase for the orbit $\Gamma $.\ It is
interesting to note that in terms of the usual covariant momentum (Peierls
substitution) we have instead of Eq.\ $\left( \ref{pipi}\right) $ the
condition $\oint \Pi _{x}d\Pi _{y}=2\mathcal{\pi }\hbar eB\left(
n+1/2\right) .$ The integration in Eq. $\left( \ref{pipi}\right) $ defines
the cross-sectional area $S_{0}(\varepsilon ,\pi _{z})$ of the orbit $\Gamma 
$ which is the intersection of the constant energy surface $\varepsilon
\left( \pi \right) =$const and the plane $\pi _{z}=$const. Therefore the
condition Eq. $\left( \ref{pipi}\right) $ implicitly determines the energy
levels $\varepsilon _{n}\left( \pi _{z}\right) .$ Computing now the
cross-sectional area $S_{0}(\mathcal{E},\pi _{z})=S_{0}(\varepsilon -e\hbar
M_{z}B,\pi _{z})\approx S_{0}(\varepsilon ,\pi _{z})+dS,$ with $dS=\oint
d\kappa d\mathbf{\pi }_{\perp }$ the area of the annulus between the energy
surface $\varepsilon =$const and the surface $\varepsilon +d\varepsilon $
with $d\varepsilon =-e\hbar M_{z}B,$ and where $d\kappa =\sqrt{d\pi
_{x}^{2}+d\pi _{y}^{2}}$ is an elementary length of the $\pi $ orbit. Then,
as $dS$ can be written $dS=\oint \frac{d\varepsilon d\kappa }{\left\vert
\partial \varepsilon /\partial \mathbf{\pi }_{\perp }\right\vert }=-e\hbar
B\oint \frac{M_{z}d\kappa }{\left\vert \partial \varepsilon /\partial 
\mathbf{\pi }_{\perp }\right\vert }$ where the integral is taken over the
orbit $\Gamma $, one has finally 
\begin{equation}
S_{0}(\mathcal{E},\pi _{z})=2\mathcal{\pi }\hbar eB\left( n+\frac{1}{2}-\phi
_{B}-\frac{1}{2\mathcal{\pi }}\oint \frac{M_{z}\left( \mathbf{\pi }\right)
d\kappa }{\left\vert \partial \varepsilon /\partial \mathbf{\pi }_{\perp
}\right\vert }\right)  \label{S0}
\end{equation}%
It is common to write $S_{0}(\mathcal{E},\pi _{z})=2\mathcal{\pi }\hbar
eB\left( n+\gamma \right) $ defining thus the coefficient $\gamma -\frac{1}{2%
}=-\phi _{B}-\frac{1}{2\mathcal{\pi }}\oint \frac{M_{z}d\kappa }{\left\vert
\partial \varepsilon /\partial \mathbf{\pi }_{\perp }\right\vert }.$ This
coefficient can also be written in a different form 
\begin{equation}
\gamma -\frac{1}{2}=-\frac{1}{2\mathcal{\pi }}\oint \frac{\left[ \widetilde{%
\mathbf{v}}\times \emph{A}+\mathbf{M}\right] _{z}d\kappa }{\left\vert
\partial \varepsilon /\partial \mathbf{\pi }_{\perp }\right\vert }
\label{gamma}
\end{equation}%
with $\widetilde{\mathbf{v}}\mathbf{\equiv }\partial \varepsilon /\partial 
\mathbf{\pi .}$ Eq. $\left( \ref{gamma}\right) $ is a generalization of a
previous result found by Roth \cite{Roth} and Fal'kovskii \cite{KALKOVSKII},
in the specific context of Bloch electrons in a magnetic field. The
connection with Berry's phase was seen later by Mikitik and Sharlai \cite%
{MIKITIK}. In both \cite{Roth} \ and \cite{MIKITIK}, the term $\left[ 
\widetilde{\mathbf{v}}\times \emph{A}+\mathbf{M}\right] $ was written as $%
\frac{1}{2}\mathcal{P}_{n}\left[ \left( \frac{\mathbf{\Pi }}{m}+\mathbf{v}%
\right) \times \mathcal{A}\right] $ where $\mathbf{v=}\frac{\mathbf{\Pi }}{m}%
+\hbar \overset{\cdot }{\mathcal{A}}$ is the velocity operator before
projection on a band, and $\Pi =m\overset{\cdot }{\mathbf{R}},$ a relation
valid only for a Hamiltonian whose kinetic energy is $\Pi ^{2}/2m.$
Therefore Eq. $\left( \ref{S0}\right) $ is more general and has a broader
field of application, as it is a general result which applies for any kind
of single quantum particle system in a magnetic field, including Bloch and
Dirac electrons. Importantly the derivation provided here is new, and it
turns out to be the result of the generalized Peierls substitution in the
Bohr-Sommerfeld condition.

\textit{Bloch electron. }In a crystal, the Berry gauge $\emph{A}\left( 
\mathbf{k}\right) $ is Abelian (a scalar operator), written in terms of the
periodic part of the Bloch wave $\left\vert u_{n}(\mathbf{k})\right\rangle $
as $\emph{A}\left( \mathbf{k}\right) \emph{=}i\left\langle u_{n}(\mathbf{k}%
)\right\vert \partial _{\mathbf{k}}\left\vert u_{n}(\mathbf{k})\right\rangle 
$, where $\mathbf{k}$ is the generalized covariant pseudo momentum ($\mathbf{%
k=\pi /\hbar }$). Application of Eq. $\left( \ref{S0}\right) $ for electron
trajectories in a crystal with time reversal and spatial inversion symmetry,
where it is expected that, both $\mathbf{\Theta }$ and $\mathbf{M}$ vanish
in the Brillouin zone, has been studied by Mikitik and Sharlai \cite{MIKITIK}%
. But these authors also pointed out the fact that the Berry's phase is non
zero when the electron orbit surrounds the band-contact line of a metal,
actually $\phi _{B}=\pm 1/2$. Consequently, $\gamma =0$ in this case,
instead of the previously supposed constant value $\gamma =1/2$ which is
commonly used in describing oscillation effect in metals. As these authors
mentioned, measurements of $\gamma $ can allows the detection of band
contact lines.

As a simple application of Eq. $\left( \ref{S0}\right) $ consider a crystal
with time reversal and spatial inversion symmetry, and where the Fermi
surface is an ellipsoid of revolution characterized by two effective masses,
a transverse $m_{\perp }$ and a longitudinal $m_{l}$ one. The energy levels
can easily be deduced. Indeed $\mathcal{E}=\hbar ^{2}\left( \frac{\mathbf{k}%
_{\perp }^{2}}{2m_{\perp }}+\frac{K_{z}^{2}}{2m_{l}}\right) $ and the
cross-sectional area $S_{0}(\mathcal{E},K_{z})$ is a disc of radius square $%
\mathbf{k}_{\perp }^{2}=2m_{\perp }\left( \mathcal{E}\mathbf{/}\hbar ^{2}-%
\frac{K_{z}^{2}}{2m_{l}}\right) $ so that the energy levels are $\mathcal{E}%
_{n}=\frac{eB\hbar }{m_{\perp }}\left( n+\frac{1}{2}\right) +\frac{\hbar
^{2}K_{z}^{2}}{2m_{l}}$ which actually coincide with the exact ones because
the energy levels of an harmonic oscillator keep their form at large $n$.

\textit{Dirac electron.} Let us consider\textit{\ }the Dirac Hamiltonian $H=%
\mathbf{\alpha .\Pi }+\beta m$ in the presence of an uniform magnetic field,
with $\alpha $ and $\beta $ the usual $\left( 4\times 4\right) $ Dirac
matrices. The semiclassical block-diagonalization followed by a projection
on, say, the positive energy subspace, leads to the ($2\times 2$) matrix
valued energy operator $\mathcal{E=}\varepsilon -e\hbar \mathbf{M}.\mathbf{B}
$ where $\varepsilon =\sqrt{\mathbf{\pi }^{2}+m^{2}}$ ($c=1$) and the
magnetic moment is given by $\mathbf{M}=\frac{\mathbf{\sigma }}{2\varepsilon 
}-\frac{\mathbf{L}}{\varepsilon },$ with $\mathbf{L=\pi \times }\emph{A}$
representing the intrinsic orbital angular momentum \cite{BLIOKH}\cite%
{PIERRESC}. It turns out that for Dirac, the magnetic moment can also be
expressed as $\mathbf{M=}\varepsilon \mathbf{\Theta }$, with the curvature
vector given by the matrix \cite{BLIOKH}\cite{PIERRESC} 
\begin{equation*}
\mathbf{\Theta }\left( \mathbf{\pi }\right) =-\frac{1}{2\varepsilon ^{3}}%
\left[ m\mathbf{\sigma }+\frac{\left( \mathbf{\sigma .\pi }\right) \mathbf{%
\pi }}{\varepsilon +m}\right] 
\end{equation*}%
with $\mathbf{\sigma }$ the Pauli matrices. Berry's connection is defined as 
$\emph{A=}i\left\langle +,\mathbf{\pi }\right\vert \partial _{\mathbf{\pi }%
}\left\vert +,\mathbf{\pi }\right\rangle $ where $\left\vert +,\mathbf{\pi }%
\right\rangle $ is two components spinor of the positive energy subspace.
Consider $\mathbf{B}$ pointing in the $z$-direction so that $\pi _{z}=P_{z}=$%
const, with the goal to compute the Landau energy levels (LEL) as an
application of Eq. $\left( \ref{S0}\right) .$ As the cross-sectional area $%
S_{0}(\varepsilon ,P_{z})$ is a disc of radius square $\mathbf{\pi }_{\perp
}^{2}=\varepsilon ^{2}-m^{2}-P_{z}^{2}$, the application of Eq.\ $\left( \ref%
{S0}\right) $ consists in replacing $\varepsilon $ by $\mathcal{E}$ in $%
\mathbf{\pi }_{\perp }^{2}$ so that we have $S_{0}(\mathcal{E},P_{z})=%
\mathcal{\pi }\left( \mathcal{E}_{n}^{2}-m^{2}-P_{z}^{2}\right) $, which
yields the semiclassical quantized LEL through the relation 
\begin{equation*}
\mathcal{E}_{n}^{2}-m^{2}-P_{z}^{2}=2\hbar eB\left( n+\frac{1}{2}-\phi _{B}-%
\frac{1}{2\mathcal{\pi }}\oint \frac{M_{z}d\kappa }{\left\vert \partial
\varepsilon /\partial \mathbf{\pi }_{\perp }\right\vert }\right) 
\end{equation*}%
Now from the Berry connection $\emph{A}=\frac{\mathbf{\pi \times \sigma }}{%
2\varepsilon \left( \varepsilon +m\right) }$ we deduce the Berry's phase $%
\phi _{B}=-\frac{\tau }{2}+\tau \left( \frac{m}{2\varepsilon }+\frac{%
P_{z}^{2}}{2\varepsilon \left( \varepsilon +m\right) }\right) $ where $\tau
=\pm 1$ are the eigenvalues of the Pauli matrix $\sigma _{z}.$ Berry's phase
is the sum of a topological part $-\frac{\tau }{2}$ and a non-topological $%
\tau \left( \frac{m}{2\varepsilon }+\frac{P_{z}^{2}}{2\varepsilon \left(
\varepsilon +m\right) }\right) $ one. The contribution from the magnetic
moment yields $\frac{1}{2\mathcal{\pi }}\oint \frac{M_{z}d\kappa }{%
\left\vert \partial \varepsilon /\partial \mathbf{\pi }_{\perp }\right\vert }%
=-\tau \left( \frac{m}{2\varepsilon }+\frac{P_{z}^{2}}{2\varepsilon \left(
\varepsilon +m\right) }\right) $ a term which exactly cancels the
non-topological contribution of $\phi _{B},$ so that finally 
\begin{equation*}
\mathcal{E}_{n}=\sqrt{m^{2}+2\hbar Be\left( n+\frac{1}{2}+\frac{\tau }{2}%
\right) +P_{z}^{2}}
\end{equation*}%
It turns out in this example that the semiclassical energy quantization
coincides also with the exact result. It is usually expected that for a
massless Dirac particle the Berry's phase takes the topological value $\phi
_{B}=\pm 1/2$, as a consequence of the band degeneracy at zero momentum \cite%
{BERRY}. This is not the case here because the magnetic field lifts this
degeneracy as $P_{z}$ is not zero in, general. But it turns out that the
magnetic moment contribution exactly compensates for the non-topological
Berry's phase contribution. This cancellation can be easily understood from
the expression Eq.\ $\left( \ref{gamma}\right) $ for the coefficient $\gamma
.$ Indeed from the equality $\left[ \widetilde{\mathbf{v}}\times \emph{A}+%
\mathbf{M}\right] _{z}=\left[ \frac{\mathbf{\pi }\times \emph{A}}{%
\varepsilon }\right] _{z}+\frac{\tau }{2\varepsilon }-\frac{\mathbf{L}_{z}}{%
\varepsilon }=\frac{\tau }{2\varepsilon }$ we deduce the expected result $%
\gamma =\frac{1}{2}+\frac{\tau }{2}=0$ or $1$.

For a two-dimensional Dirac system it is therefore expected that the
magnetic moment for massless particles exactly vanishes, and that the
Berry's phase takes the topological value $\phi _{B}=\pm 1/2.$ The electron
motion in graphene is an interesting physical situation which illustrates
this assertion. Indeed, graphene is a two-dimensional carbon crystalline
honeycomb structure with inversion symmetry so that $\mathbf{M=}0$. The
hexagonal Brillouin zone has two distinct and degenerate Dirac points or
valleys (labelled by $\tau \pm 1$) where the conduction and valence bands
meet and the electronic excitations behave like massless relativistic
fermions, so that $\phi _{B}=\pm 1/2$ and consequently $\mathcal{E}_{n}=\pm 
\sqrt{2\hbar eB\left( n+\frac{1}{2}+\frac{\tau }{2}\right) }$ \cite%
{PIERREGRAPHENE}. Therefore the ground state is not degenerate as there is
only one possibility to realize it $n=0$. This result explains the peculiar
quantum Hall effect of graphene \cite{rev}.

\textit{Summary}. We have shown that a generalized Peierls substitution
including a Berry phase term must be considered for a correct semiclassical
treatment of electrons in a magnetic field. This substitution is essential
for the determination of the full semiclassical equations of motion, as well
as for the semiclassical Bohr-Sommerfeld quantization condition for energy
levels. Indeed, the substitution in the Bohr-Sommerfeld condition leads to
an expression for the cross-sectional area which in some sort generalizes
the formula found by Roth and Fal'kovskii in the context of Bloch electrons
in a crystal. Application of this formula to Dirac electrons shows the
subtle cancellation mechanism between the magnetic moment and the
non-topological part of the Berry's phase, which yields the Landau energy
levels.

\textit{Acknowledgement}. The authors acknowledge fruitful discussions with
F. Pi\'{e}chon and J.N. Fuchs. We also thank L. A. Fal'kovskii for having
drawn our attention to his own work on this subject.

\end{document}